\documentclass{article}
\usepackage{amssymb}
\usepackage{geometry}
\usepackage{eurosym}
\usepackage{amsmath}
\usepackage{graphicx}

\usepackage{geometry}
\usepackage{eurosym}
\usepackage{graphicx}
\usepackage{mathptmx}    
\usepackage{amsthm,amsmath,amssymb,latexsym,amscd}
\usepackage{tabularx}
\usepackage{mathrsfs} 
\usepackage{diagbox}

\usepackage{latexsym}
\usepackage[mathscr]{eucal}
\usepackage{mathtools}
\usepackage{float}
\restylefloat{figure}

\usepackage{mathptmx}      
\usepackage{latexsym}
\usepackage{booktabs}
\usepackage{multirow}
\usepackage{tabularx}
\usepackage{caption}

\def\bn{\begin{equation}}
\def\en{\end{equation}}

\begin{document}

\title{\textbf{Traveling wave solutions of the generalized scale-invariant analogue of the KdV equation by tanh–coth method}}

\author{O. Gonz\'alez-Gaxiola $^{1\; \ast}$, J. Ruiz de Ch\'avez $^{2}$ \\
\\
$^{1}$ Applied Mathematics and Systems Department, Universidad Aut\'onoma Metropolitana-Cuajimalpa, \\
Vasco de Quiroga 4871, 05348 Mexico City, Mexico.\\
$^{\ast}$ ogonzalez@cua.uam.mx
\\
$^2$ Department of Mathematics, Universidad Aut\'onoma Metropolitana-Iztapalapa. \\ San Rafael Atlixco 186, Col. Vicentina, Iztapalapa, 09340, Mexico City, Mexico.
}


\date{}

\maketitle

\begin{abstract}
\noindent In this work, the generalized scale-invariant analogue of the Korteweg–de Vries (gsiaKdV) equation is studied. For the first time, the tanh–coth methodology is used to find traveling wave solutions for this nonlinear  equation. The considered generalized equation is a connection between the well-known KdV equation and the recently investigated SIdV equation. The obtained results show many families of solutions for the model, indicating that this equation also shares bell-shaped solutions with KdV and SIdV, as previously documented by other researchers.  Finally, by
executing the symbolic computation, we demonstrate that the employed technique is a valuable and effective mathematical tool that can be used to solve problems that arise in the cross-disciplinary nonlinear sciences and applied mathematics.

\end{abstract}

\begin{center}
\end{center}

\section{Introduction}
\label{s:Intro}
\noindent Many fields of science and engineering depend heavily on the mathematical models presented by nonlinear partial differential equations (NPDE) to explain complex phenomena. These fields include electromagnetic waves theory, ocean dynamics, plasma physics, fluid mechanics, field theory, nonlinear optical fibers, nuclear physics,  ion acoustic waves, biological process engineering, chemical kinetics, climatological phenomena and several other mathematical physics problems. In 1895, Dutch mathematicians D. J. Korteweg and G. de Vries developed the KdV equation  in a formal manner, the KdV equation is a NPDE that models the propagation of long waves on shallow water and this simultaneously includes weak advective nonlinearity and dispersion effects given by \cite{KdV1}
\begin{equation}\label{K0}
	u_{t}+6uu_{x}+u_{xxx}=0,
\end{equation}
where $u$ is the perturbation wave function that depends on the spatial variable $x$ and on time $t$. 
It is well known that Eq. (\ref{K0}) has bell-shaped solutions of the type:
\begin{equation}\label{K01}
	u(x,t)=\frac{c}{2}sech^{2}\Big(\frac{\sqrt{c}}{2}(x-ct)\Big)
\end{equation}
where $c$ is the velocity of the wavefront \cite{KdV2}.\\
Despite its antiquity, the KdV equation is still an active area of study and research, with several articles published on the topic in recent years \cite{KdV3,KdV51,KdV6,KdV5,KdV4,KdV7,A-9,Os,Ku1,Ku2,Bis1}.\\
Among the numerous research associated to the KdV equation that have been published in the last decades, several propose to generalize and/or modify the KdV equation. In 2012, for instance, the authors of \cite{KdV3} present the modified KdV equation
\begin{equation}\label{K1}
	u_{t}+\Big(\frac{2u_{xx}}{u}\Big)u_{x}= u_{xxx}.
\end{equation}
The Eq. (\ref{K1}) is invariant under scaling of the dependent variable; is therefore referred to as SIdV here. Eq. (\ref{K1}) was discovered by surprise using computer approaches when researchers explored equations with bell-shaped solutions analogous to the KdV equation. Other research on Eq. (\ref{K1}) have been published and can be found in \cite{Re-3,Re-4,Re-41}.\\
In this article we will consider the gsiaKdV equation \cite{Re-1} which is a  nonlinear partial differential equation and whose dimensionless form is given by
\begin{equation}\label{Eq-1}
	u_{t}+\Big(3(1-\alpha)u+(1+\alpha)\frac{u_{xx}}{u}\Big)u_{x}=\gamma u_{xxx}
\end{equation}
where $u_{xxx}$ is a dispersion term, while the term $\big(3(1-\alpha)u+(1+\alpha)\frac{u_{xx}}{u}\big)$ can be seen as an advecting velocity.\\
Firstly, let us observe that,  if $\alpha=-1$ and $\gamma=-1$, then Eq. (\ref{Eq-1}) reduces to the well-known KdV equation (\ref{K0}).  Second, we can observe that  if $\alpha=1$ and $\gamma=1$ then Eq. (\ref{Eq-1}) reduces to SIdV equation (\ref{K1}).\\
Eq. (\ref{Eq-1}) was investigated in \cite{Re-5} and the authors demonstrated the existence of traveling waves of the bell and valley types. Using the tanh-coth method for the first time, the main objective of this research is to find new traveling wave type solutions for Eq. (\ref{Eq-1}) with $\alpha\neq\pm1$ and $\gamma\neq 0$.\\
\noindent In virtue of the significance of the previously described gsiaKdV equation, we will conduct a study to obtain solutions of the traveling-wave type for the first time utilizing the tanh-coth technique.  In addition, some 3D and 2D propagation profiles for the derived solutions will be discussed by selecting various parameters that describe the solution sets achieved by the used strategy. 

\section{ Brief description of the tanh-coth method}\label{Met}

\noindent The tanh-coth method originally established in \cite{Re-6,Re-7} provides a very useful methodology for finding traveling wave type solutions of NPDEs. We will explain how to implement the method in the rest of this section. \\
({\bf I}) Consider the general nonlinear PDE given by:
\begin{eqnarray}\label{E-2}
	G(u,u_{t}, u_{x}, u_{xx}, u_{xxx},\ldots)=0.
\end{eqnarray}
Using traveling wave variable change $u(x,t)=u(\xi)$ with $\xi=cx-\omega t$, Eq. (\ref{E-2}) becomes the ordinary differential equation:
\begin{eqnarray}\label{E-3}
	F(u,-\omega u_{\xi}, cu_{\xi}, c^2 u_{\xi\xi}, c^3 u_{\xi\xi\xi},\ldots)=0.
\end{eqnarray}
({\bf II}) The tanh-coth method provides the solutions for Eq (\ref{E-3})  as the finite sum
\begin{eqnarray}\label{E-4}
	u(\xi)=S(Y)=\sum_{i=0}^{M}a_{i}Y^{i}(\xi)+\sum_{i=1}^{M}b_{i}Y^{-i}(\xi),
\end{eqnarray}
where the coefficients $a_i$ and $b_i$ are constants to be determined and $Y$ is a new dependent variable introduced by the method and is given by
\begin{eqnarray}\label{E-5}
	Y=\tanh(\xi).
\end{eqnarray}
The introduction of this new dependent variable implies that:
\begin{equation*}
	u_{\xi}=(1-Y^2)\frac{dS}{dY},
\end{equation*}
\begin{equation}\label{e}
	u_{\xi\xi}=-2Y(1-Y^2)\frac{dS}{dY}+(1-Y^2)^{2}\frac{d^{2}S}{dY^{2}},
\end{equation}
\begin{equation*}
	u_{\xi\xi\xi}=2(1-Y^2)(3Y^2 -1)\frac{dS}{dY}-6Y(1-Y^2)^{2}\frac{d^{2}S}{dY^{2}}+(1-Y^2)^{3}\frac{d^{3}S}{dY^{3}},
\end{equation*}
The subsequent derivatives can be computed in a similar way. \\
({\bf III}) To determine the upper limit $M$ of the sum in Eq. (\ref{E-4}), the linear terms of highest order in the resulting equation with the highest order nonlinear terms are balanced.\\
({\bf IV}) We consider $u(\xi)$ given in (\ref{E-4}) and the necessary derivatives $u_\xi$, $u_{\xi\xi}$, $u_{\xi\xi\xi}$,..., which can be calculated as in (\ref{e}), to substitute in the ordinary differential equation (\ref{E-3}) and thus we will obtain the polynomial equation:
\begin{eqnarray}\label{E-7}
	P[Y]=0.
\end{eqnarray}
({\bf V}) We select all the terms that have the same algebraic power of $Y$ from the polynomial equation (\ref{E-7}), we set them equal to zero and obtain a nonlinear system of algebraic equations with the set of unknown parameters $\{a_{0}, \ldots, a_{M}, b_{1},\ldots,b_{M}, c,\omega\}$. Using software such as  Mathematica, we can execute symbolic calculations to solve the algebraic equations with the natural restrictions of the mathematical model.\\
({\bf VI}) Finally, having obtained the coefficients $\{a_{0}, \ldots, a_{M}, b_{1},\ldots,b_{M}, c,\omega\}$ and considering the equality (\ref{E-4}) one can obtain the exact solutions to Eq. (\ref{E-2}).

\section{Utilization of the tanh-coth methodology }
\noindent Using the change of variable $\xi=cx-\omega t$, Eq. (\ref{Eq-1}) is converted into the ordinary differential equation: 
\begin{eqnarray}\label{e1}
	-\omega cuu_{\xi}+3c(1-\alpha)u^{2}u_{\xi}+c^{3}(1+\alpha)u_{\xi}u_{\xi\xi}-\gamma c^{3}uu_{\xi\xi\xi}=0.
\end{eqnarray}
Integrating once with respect to $\xi$ and considering the constants of integration as null, we obtain
\begin{eqnarray}\label{e2}
	-\omega cu^{2}+2c(1-\alpha)u^{3}+c^{3}(1+\alpha+\gamma)u_{\xi}^{2}-2\gamma c^{3}uu_{\xi\xi}=0.
\end{eqnarray}
Then using the characteristic variable change of the method, i.e., $Y=\tanh(\xi)$ and considering Eq. (\ref{E-4}), the last differential equation is rewritten as
\begin{equation}\label{e3}
	-\omega cS^{2}+2c(1-\alpha)S^{3}+c^{3}(1+\alpha+\gamma)(1-Y^2)^{2}\Big(\frac{dS}{dY}\Big)^{2}-2\gamma c^{3}S\Big[(1-Y^2)^{2}\frac{d^{2}S}{dY^{2}}-2Y(1-Y^2)\frac{dS}{dY}\Big]=0.
\end{equation}
Balancing $S^3$ with $S\cdot \frac{d^{2}S}{dY^{2}}$ gives $M=2$.
Consequently, the tanh-coth  technique enables the use of the finite sum 
\begin{eqnarray}\label{e4}
	u(\xi)=S(Y)=a_{0}+a_{1}Y+a_{2}Y^2+b_{1}Y^{-1}+b_{2}Y^{-2}.
\end{eqnarray}
Substituting (\ref{e4}) with their respective derivatives into (\ref{e3}) and collecting all terms with equal power of $Y$, after some algebraic simplification, we obtain the following nonlinear  system of algebraic equations:
\begin{gather*} 
	12 a_1 b_1 \gamma  c^3+4 a_0 b_2 \gamma  c^3+48 a_2 b_2 \gamma  c^3+4 a_1 b_1 c^3 \alpha +16 a_2 b_2 c^3 \alpha+4 a_1 b_1 c^3+16 a_2 b_2 c^3-6 a_2 b_1^2 c \alpha \\ -12 a_0 a_1 b_1 c \alpha-6 a_1^2 b_2 c \alpha -12 a_0 a_2 b_2 c \alpha  -2 a_1 b_1 c \omega -2 a_2 b_2 c \omega +6 a_2 b_1^2 c+12 a_0 a_1 b_1 c+6 a_1^2 b_2 c+12 a_0 a_2 b_2 c\\+a_1^2 \gamma  c^3 -4 a_0 a_2 \gamma  c^3 +a_1^2 c^3 \alpha +a_1^2 c^3-2 a_0^3 c \alpha -a_0^2 c \omega +2 a_0^3 c+5 b_1^2 \gamma  c^3-8 b_2^2 \gamma  c^3+b_1^2 c^3 \alpha +b_1^2 c^3=0,
\end{gather*} 
\begin{gather*} 
	-7 a_1^2 \gamma  c^3+8 a_2^2 \gamma  c^3-20 a_0 a_2 \gamma  c^3 +a_1^2 c^3 \alpha -8 a_2^2 c^3 \alpha +a_1^2 c^3-8 a_2^2 c^3-6 a_0 a_2^2 c \alpha -6 a_1^2 a_2 c \alpha -a_2^2 c \omega\\+6 a_0 a_2^2 c+6 a_1^2 a_2 c=0,
\end{gather*} 
\begin{gather*} 
	4 a_1^2 \gamma  c^3-16 a_2^2 \gamma  c^3+8 a_0 a_2 \gamma  c^3+4 a_2^2 c^3 \alpha +4 a_2^2 c^3-2 a_2^3 c \alpha +2 a_2^3 \gamma=0,
\end{gather*} 
\begin{gather*} 
	4 a_2 b_1 \gamma\alpha+4 a_0 a_1 \gamma  c^3-24 a_1 a_2 \gamma  c^3+4 a_1 a_2 c^3 \alpha +4 a_1 a_2 c^3-6 a_1 a_2^2 c \alpha +6 a_1 a_2^2 \omega=0,
\end{gather*} 
\begin{gather*} 
	-4 a_0 b_1 \gamma  c^3-20 a_2 b_1 \gamma  c^3-4 a_1 b_2 \gamma  c^3-4 a_2 b_1 c^3 \alpha -4 a_2 b_1 c^3-6 a_2^2 b_1 c \alpha +6 a_2^2 b_1 c-8 a_0 a_1 \gamma  c^3+12 a_1 a_2 \gamma  c^3 \\-8 a_1 a_2 c^3 \alpha -8 a_1 a_2 c^3-2 a_1^3 c \alpha -12 a_0 a_1 a_2 c \alpha -2 a_1 a_2 c \omega +2 a_1^3 c+12 a_0 a_1 a_2 c=0,
\end{gather*} 
\begin{gather*} 
	-6 a_1 b_1 \gamma  c^3-8 a_0 b_2 \gamma  c^3-24 a_2 b_2 \gamma  c^3-2 a_1 b_1 c^3 \alpha -8 a_2 b_2 c^3 \alpha -2 a_1 b_1 c^3-8 a_2 b_2 c^3-12 a_1 a_2 b_1 c \alpha -6 a_2^2 b_2 c\alpha\\ +12 a_1 a_2 b_1 c+6 a_2^2 b_2 c+2 a_1^2 \gamma  c^3+16 a_0 a_2 \gamma  c^3-2 a_1^2 c^3 \alpha +4 a_2^2 c^3 \alpha -2 a_1^2 c^3+4 a_2^2 c^3-6 a_0 a_1^2 c \alpha -6 a_0^2 a_2 c\alpha\\ -a_1^2 c \omega -2 a_0 a_2 c \omega +6 a_0 a_1^2 c+6 a_0^2 a_2 c-4 b_1^2 \gamma  c^3=0,
\end{gather*} 
\begin{gather*} 
	4 a_0 b_1 \gamma  c^3+28 a_2 b_1 \gamma  c^3-8 a_1 b_2 \gamma  c^3+8 a_2 b_1 c^3 \alpha -4 a_1 b_2 c^3 \alpha +8 a_2 b_1 c^3-4 a_1 b_2 c^3-6 a_1^2 b_1 c \alpha -12 a_0 a_2 b_1 c \alpha \\ -12 a_1 a_2 b_2 c \alpha -2 a_2 b_1 c \omega +6 a_1^2 b_1 c+12 a_0 a_2 b_1 c+12 a_1 a_2 b_2 c+4 a_0 a_1 \gamma  c^3+4 a_1 a_2 c^3 \alpha +4 a_1 a_2 c^3-6 a_0^2 a_1 c \alpha \\ -2 a_0 a_1 c \omega +6 a_0^2 a_1 c-12 b_1 b_2 \gamma  c^3=0,
\end{gather*} 
\begin{gather*} 
	4 a_0 b_1 \gamma  c^3-12 a_2 b_1 \gamma  c^3+28 a_1 b_2 \gamma  c^3-4 a_2 b_1 c^3 \alpha+8 a_1 b_2 c^3 \alpha -4 a_2 b_1 c^3+8 a_1 b_2 c^3-6 a_1 b_1^2 c \alpha -6 a_0^2 b_1 c \alpha\\ -12 a_0 a_1 b_2 c \alpha -12 a_2 b_1 b_2 c \alpha -2 a_0 b_1 c \omega -2 a_1 b_2 c \omega +6 a_1 b_1^2 c+6 a_0^2 b_1 c+12 a_0 a_1 b_2 c+12 a_2 b_1 b_2 c\\+12 b_1 b_2 \gamma  c^3+4 b_1 b_2 c^3 \alpha +4 b_1 b_2 c^3=0,
\end{gather*} 
\begin{gather*} 
	-12 a_0 b_2 \gamma  c^3-6 a_0 b_2^2 c \alpha+6 a_0 b_2^2 c-3 b_1^2 \gamma  c^3+8 b_2^2 \gamma  c^3+b_1^2 c^3 \alpha-8 b_2^2 c^3 \alpha +b_1^2 c^3-8 b_2^2 c^3\\ -6 b_1^2 b_2 c \alpha -b_2^2 c \omega +6 b_1^2 b_2 c=0,
\end{gather*} 
\begin{gather*} 
	-4 a_0 b_1 \gamma  c^3-16 a_1 b_2 \gamma  c^3-4 a_1 b_2 c^3 \alpha -4 a_1 b_2 c^3-6 a_1 b_2^2 c \alpha -12 a_0 b_1 b_2 c \alpha +6 a_1 b_2^2 c+12 a_0 b_1 b_2 c\\ +12 b_1 b_2 \gamma  c^3-8 b_1 b_2 c^3 \alpha -8 b_1 b_2 c^3-2 b_1^3 c \alpha -2 b_1 b_2 c \omega +2 b_1^3 c=0,
\end{gather*} 
\begin{gather*} 
	-6 a_1 b_1 \gamma  c^3+16 a_0 b_2 \gamma  c^3-24 a_2 b_2 \gamma  c^3-2 a_1 b_1 c^3 \alpha -8 a_2 b_2 c^3 \alpha -2 a_1 b_1 c^3-8 a_2 b_2 c^3-6 a_0 b_1^2 c \alpha -6 a_2 b_2^2 c \alpha\\  -6 a_0^2 b_2 c \alpha -12 a_1 b_1 b_2 c \alpha -2 a_0 b_2 c \omega +6 a_0 b_1^2 c+6 a_2 b_2^2 c+6 a_0^2 b_2 c+12 a_1 b_1 b_2 c+2 b_1^2 \gamma  c^3+8 b_2^2 \gamma  c^3-2 b_1^2 c^3 \alpha\\ +4 b_2^2 c^3 \alpha -2 b_1^2 c^3+4 b_2^2 c^3-b_1^2 c \omega =0,
\end{gather*} 
\begin{gather*} 
	-12 b_1 b_2 \gamma  c^3+4 a_1 b_1 b_2 c^3 \alpha +4 b_1 b_2 c^3-6\omega b_1 b_2^2  \alpha +6a_0 b_1 b_2^2 c=0,
\end{gather*} 
\begin{gather*} 
	-8 b_2^2 \gamma  c^3 +4 a_0 a_1 \gamma \omega+4 b_2^2 c^3 \alpha +4a_2 b_2^2 c^3-2 b_2^3 c \alpha +2 b_2^3 c=0.
\end{gather*} 
\noindent Using the well-known {\it Mathematica}  software to solve the above system, we find the following families of solutions: \\
\noindent {\bf Family 1:} For $\alpha\neq 1 $ and  $c\neq 0$:
$$
a_0=a_0, \;\, a_1= 0,\,\; a_2=a_2,\, \; b_1= 0,\; \; b_2= -\frac{2 c^2 (2 \gamma -\alpha-1)}{\alpha\ -1}, \; \omega\neq 0.
$$
Substituting the obtained parameters into the general solution (\ref{e4}), we obtain the following family of solutions
\begin{equation}\label{S-1}
	u_{1}(x,t)=a_0+a_2\tanh^{2}(cx-\omega t)-\frac{2 c^2 (2 \gamma -\alpha-1)}{\alpha\ -1}\coth^{2}(cx-\omega t).
\end{equation}
\noindent {\bf Family 2:} For $\alpha\neq \pm 1$ and $c\neq 0$:
$$
a_0=0, \, a_1= a_1,\, a_2=-\frac{2 c^2 (4 \gamma ^2-\alpha ^2-2 \alpha-1)}{5 ( \alpha ^2-1)},\, b_1= 0, \; b_2= -\frac{2 c^2 (2 \gamma -\alpha-1)}{\alpha -1}, \; \omega\neq 0.
$$
Therefore, proceeding as in the previous case, the set of solutions for this family is provided by
\begin{equation}\label{S-2}
	u_{2}(x,t)=a_1\tanh(cx-\omega t)-\frac{2 c^2 (4 \gamma ^2-\alpha ^2-2 \alpha-1)}{5( \alpha ^2-1)}\tanh^{2}(cx-\omega t)-\frac{2 c^2 (2 \gamma -\alpha-1)}{\alpha\ -1}\coth^{2}(cx-\omega t).
\end{equation}
\noindent {\bf Family 3:} For $\alpha\neq  1$ and $c\neq 0$:
$$
a_0=a_0\neq 0, \, a_1= a_1,\, a_2=\frac{3}{2}a_1,\, b_1= 0,\; b_2= -\frac{2 c^2 (2 \gamma -\alpha -1)}{ \alpha-1 }, 
$$
$$\omega=\frac{8 a_0 \gamma  c^2 \alpha -8 a_0 \gamma  c^2+6 a_0^2 \alpha-16 \gamma ^2 c^4+4 c^4 \alpha ^2 +4 c^4}{a_0 (\alpha -1)}.$$
Therefore, proceeding as in the previous cases, the set of solutions for this family is provided by
\begin{equation}\label{S-3}
	u_{3}(x,t)=a_0+a_1\tanh(cx-\omega t)+\frac{3}{2}a_1\tanh^{2}(cx-\omega t)-\frac{2 c^2 (2 \gamma -\alpha-1)}{\alpha\ -1}\coth^{2}(cx-\omega t).
\end{equation}
\noindent {\bf Family 4:} For $\alpha\neq \pm 1$ and $c\neq 0$ :
$$
a_0=a_0, \, a_1= a_1,\, a_2=\frac{3}{4}a_1,\, b_1= 0,\; b_2= -\frac{2 c^2 (2 \gamma -\alpha -1)}{\alpha -1},
\; \omega=\frac{3\gamma c^{2}-3(\alpha+1)-7\gamma}{\alpha^2 -1}. $$
Therefore, proceeding as in the previous cases, the set of solutions for this family is provided by
\begin{equation}\label{S-4}
	u_{4}(x,t)=a_0+a_1\tanh(cx-\omega t)+\frac{3}{2}a_1\tanh^{2}(cx-\omega t)-\frac{2 c^2 (2 \gamma -\alpha-1)}{\alpha\ -1}\coth^{2}(cx-\omega t).
\end{equation}
\noindent {\bf Family 5:} For $\alpha\neq \pm 1$ and $c\neq 0$:
$$
a_0=a_0, \, a_1= a_1, $$ $$ a_2=\frac{16 \gamma  c^2 \alpha ^2-16 a_0 \gamma  c^2-8 a_0 c^2 \alpha ^3-8 a_0 c^2 \alpha ^2+8 a_0 c^2 \alpha +8 a_0 c^2-3 a_0^2 \alpha ^3}{10 c^2 \left(\alpha ^2-1\right) },$$ $$ b_1= 0,\; \; b_2= -\frac{2 c^2 (2 \gamma -\alpha -1)}{\alpha-1},\;\; 
\omega=\frac{ \gamma^{2}(a_0^{3}-4a_1 \gamma  c^2-2 c^2)}{1-\alpha^2 }.$$
Therefore, proceeding as in the previous cases, the set of solutions for this family is provided by
\begin{equation} \label{S-5}
	\begin{split}
		u_{5}(x,t) & = \Big(\frac{16 \gamma  c^2 \alpha ^2-16 a_0 \gamma  c^2-8 a_0 c^2 \alpha ^3-8 a_0 c^2 \alpha ^2+8 a_0 c^2 \alpha +8 a_0 c^2-3 a_0^2 \alpha ^3}{10 c^2 \left(\alpha ^2-1\right) (2 \gamma -\alpha-1)}\Big)\tanh^{2}(cx-\omega t)\\&
		+a_0 +a_1\tanh(cx-\omega t)-\frac{2 c^2 (2 \gamma -\alpha-1)}{\alpha\ -1}\coth^{2}(cx-\omega t).
	\end{split}
\end{equation}
\noindent {\bf Family 6:} For $\alpha\neq\pm 1$, $c\neq 0$ and $2 \gamma -\alpha -1\neq 0$:
$$ a_0=a_0, \,\; a_1=\frac{20 a_0 \gamma  c^2 (\alpha -1) (2 \gamma -\alpha-1)+b_1^2 (\alpha -1)^2}{{16 c^4 (2 \gamma -\alpha -1)^2}}, \; a_2=a_2,\;\; b_1=b_1, $$ $$  b_2=-\frac{2 c^2 (2 \gamma -\alpha -1)}{\alpha -1},\;
\omega=\frac{b_1^2 (a_0\gamma -3 (\alpha +1))}{4 c^2 (\alpha^2 -1)^2}+8 c^2 (\gamma -\alpha -1).$$
Therefore, proceeding as in the previous cases, the set of solutions for this family is provided by
\begin{equation} \label{S-6}
	\begin{split}
		u_{6}(x,t) & = a_{0}+\Big(\frac{20 a_0 \gamma  c^2 (\alpha -1) (2 \gamma -\alpha-1)+b_1^2 (\alpha -1)^2}{{16 c^4 (2 \gamma -\alpha -1)^2}}\Big)\tanh(cx-\omega t)+a_{2}\tanh^{2}(cx-\omega t)\\ &+b_{1}\coth(cx-\omega t)-\frac{2 c^2 (2 \gamma -\alpha -1)}{\alpha -1}\coth^{2}(cx-\omega t).
	\end{split}
\end{equation}
\noindent {\bf Family 7:} For $\alpha\neq 1$ and $c\neq 0$:
$$
a_0=a_0,\;\, a_1= a_1, \;\; a_2=\frac{3}{2}a_1,\;\, b_1=0,\;\; b_2= -\frac{2 c^2 (2 \gamma -\alpha -1)}{\alpha -1},\;\, \omega= \left(a_0a_2+3\gamma-2\alpha^2\right).
$$
Therefore, proceeding as in the previous cases, the set of solutions for this family is provided by
\begin{equation}\label{S-7}
	u_{7}(x,t)=a_0+a_1\tanh(cx-\omega t)+\frac{3}{2}a_1\tanh^{2}(cx-\omega t)-\frac{2 c^2 (2 \gamma -\alpha-1)}{\alpha\ -1}\coth^{2}(cx-\omega t).
\end{equation}
\noindent {\bf Family 8:} For $\alpha\neq -1$, $c\neq 0$  and $3 \gamma +1\neq 0$:
$$
a_0=a_0, \;\, a_1= a_1, \;\; a_2=-\frac{3 (c\gamma^2-\alpha+3)}{2 c^2 (\alpha +1)(3 \gamma +1)},\;\; b_1= b_1,\; b_2= 0, \; \; \omega=\omega.
$$
Therefore, proceeding as in the previous cases, the set of solutions for this family is provided by
\begin{equation}\label{S-8}
	u_{8}(x,t)=a_0+a_1\tanh(cx-\omega t)-\frac{3 (c\gamma^2-\alpha+3)}{2 c^2 (\alpha +1)(3 \gamma +1)}\tanh^{2}(cx-\omega t)+b_1 \coth(cx-\omega t).
\end{equation}
\noindent {\bf Family 9:} For $\alpha\neq  1$, $c\neq 0$ and $ \gamma \neq 0$:
$$ a_0=\frac{8 a_2^2 c^2 \gamma -2 a_1^2 c^2 \gamma -2 a_2^2 c^2 \alpha -2 a_2^2 c^2+a_2^3 \alpha -a_2^3}{4 a_2 c^2 \gamma }, \; \; a_1=a_1,\;\; a_2=a_2\neq 0, $$
$$b_1=b_1,\;\; b_2=-\frac{2 c^2 (3 \gamma -\alpha-1)}{3 (\alpha -1)},\;\; \omega=\omega.$$
Therefore, proceeding as in the previous cases, the set of solutions for this family is provided by
\begin{equation} \label{S-9}
	\begin{split}
		u_{9}(x,t) & = \frac{8 a_2^2 c^2 \gamma -2 a_1^2 c^2 \gamma -2 a_2^2 c^2 \alpha -2 a_2^2 c^2+a_2^3 \alpha -a_2^3}{4 a_2 c^2 \gamma }+a_1 \tanh(cx-\omega t)+a_{2}\tanh^{2}(cx-\omega t)\\ &+b_{1}\coth(cx-\omega t)-\frac{2 c^2 (3 \gamma -\alpha-1)}{3 (\alpha -1)}\coth^{2}(cx-\omega t).
	\end{split}
\end{equation}
\noindent {\bf Family 10:} For $\alpha\neq  1$, $c\neq 0$  and $12 \gamma -5 \alpha -5\neq 0$:
$$ a_0=a_0, \; \; a_1=a_1,\;\; a_2=-\frac{2 c^2 (4 \gamma ^2-\alpha ^2-2 \alpha -1)}{(\alpha -1) (12 \gamma -5 \alpha -5)}, \;
b_1=0,\; b_2=-\frac{2 c^2a_{1} (2 \gamma -\alpha -1)}{\alpha -1},\;\;  \omega=\omega.$$
Therefore, proceeding as in the previous cases, the set of solutions for this family is provided by
\begin{equation} \label{S-10}
	\begin{split}
		u_{10}(x,t)  &= a_{0}+a_{1}\tanh(cx-\omega t)-\frac{2 c^2 (4 \gamma ^2-\alpha ^2-2 \alpha -1)}{(\alpha -1) (12 \gamma -5 \alpha -5)}\tanh^{2}(cx-\omega t)\\ & -\frac{2 c^2 a_{1}(2 \gamma -\alpha -1)}{\alpha -1}\coth^{2}(cx-\omega t).
	\end{split}
\end{equation}
\noindent {\bf Family 11:} For $\alpha\neq  1$, $c\neq 0$  and $ \gamma \neq 0$:
$$ a_0=\frac{a_2 \left(a_2 \alpha -a_2+8 \gamma  c^2-2 c^2 \alpha -2 c^2\right)}{4 c^2 \gamma }, \; \; a_1=0,\;\; a_2=a_2, $$ $$ b_1=0,\; b_2=-\frac{2 c^2 (3 \gamma -\alpha-1)}{3 (\alpha -1)},\; \omega=\omega.$$
Therefore, proceeding as in the previous cases, the set of solutions for this family is provided by
\begin{equation} \label{S-11}
	u_{11}(x,t)  = \frac{a_2 \left(a_2 \alpha -a_2+8 \gamma  c^2-2 c^2 \alpha -2 c^2\right)}{4 c^2 \gamma }+a_2 \tanh^{2}(cx-\omega t)-\frac{2 c^2 (3 \gamma -\alpha-1)}{3 (\alpha -1)}\coth^{2}(cx-\omega t).
\end{equation}
\noindent {\bf Family 12:}  For $\alpha\neq  1$, $c\neq 0$  and $ \gamma \neq 0$:
$$ a_0=\frac{b_1^2-b_1^2 \alpha }{2 c^2 \gamma }, \; \; a_1=-\frac{ \sqrt{2 a_2 c^2 (4 \gamma -\alpha -1)+a_2^2 (\alpha -1)}}{\sqrt{2} c \sqrt{\gamma }},$$
$$a_2=a_2, \;\; b_1=b_1,\;\;b_2=0,\;\;  \omega=\omega.$$
Therefore, proceeding as in the previous cases, the set of solutions for this family is provided by
\begin{equation} \label{S-12}
	\begin{split}
		u_{12}(x,t) & =\frac{b_1^2-b_1^2 \alpha }{2 c^2 \gamma }-\frac{ \sqrt{2 a_2 c^2 (4 \gamma -\alpha -1)+a_2^2 (\alpha -1)}}{\sqrt{2} c \sqrt{\gamma }}\tanh(cx-\omega t)\\ &+a_{2}\tanh^{2}(cx-\omega t)+b_{1}\coth(cx-\omega t).
	\end{split}
\end{equation}
\noindent {\bf Family 13:} For $ \gamma \neq 0$ and $c\neq 0$:
$$ a_0=\frac{8 a_2^2 c^2 \gamma -2 a_1^2 c^2 \gamma -2 a_2^2 c^2 \alpha -2 a_2^2 c^2+a_2^3 \alpha -a_2^3}{4 a_2 c^2 \gamma }, \; \; a_1=a_1,\;\; a_2=a_2\neq 0,$$
$$b_1=\frac{a_1 (16 a_2^2 c^2 \gamma +2 a_1^2 c^2 \gamma -2 a_2^2 c^2 \alpha -2 a_2^2 c^2+5 a_2^3 \alpha -5 a_2^3)}{4 a_2^2 c^2 \gamma },\;\;b_2=0,\; \omega=\omega.$$
Therefore, proceeding as in the previous cases, the set of solutions for this family is provided by
\begin{equation} \label{S-13}
	\begin{split}
		u_{13}(x,t) & =\frac{8 a_2^2 c^2 \gamma -2 a_1^2 c^2 \gamma -2 a_2^2 c^2 \alpha -2 a_2^2 c^2+a_2^3 \alpha -a_2^3}{4 a_2 c^2 \gamma }+a_1 \tanh(cx-\omega t)+a_{2}\tanh^{2}(cx-\omega t)\\ &+\frac{a_1 (16 a_2^2 c^2 \gamma +2 a_1^2 c^2 \gamma -2 a_2^2 c^2 \alpha -2 a_2^2 c^2+5 a_2^3 \alpha -5 a_2^3)}{4 a_2^2 c^2 \gamma }\coth(cx-\omega t).
	\end{split}
\end{equation}
\noindent {\bf Family 14:} For $\alpha\neq1$, $c\neq 0$  and $ \gamma \neq 0$:
$$ a_0=-\frac{3b_1\alpha+c^4}{c^2\gamma(\alpha -1)}, \; \; a_1=0,\;\; a_2=0,\;\;
b_1=b_1,\;\;b_2=-\frac{2 c^2 (2 \gamma -\alpha -1)}{\alpha -1},\; \omega=\omega.$$
Therefore, proceeding as in the previous cases, the set of solutions for this family is provided by
\begin{equation} \label{S-14}
	u_{14}(x,t)  = -\frac{3b_1\alpha+c^4}{c^2\gamma(\alpha -1)}+b_1 \coth(cx-\omega t)-\frac{2 c^2 (2 \gamma -\alpha -1)}{\alpha -1}\coth^{2}(cx-\omega t).
\end{equation}
\noindent {\bf Family 15:} For $\alpha\neq 1$, $c\neq 0$  and $9 \gamma -7\alpha -7\neq 0$:
$$ a_0=0, \; \; a_1=0,\;\; a_2=\frac{2 c^2 (6 \gamma ^2-\gamma  \alpha -\gamma -\alpha ^2-2 \alpha -1)}{(\alpha -1) (9 \gamma -7 \alpha -7)},$$
$$ b_1=b_1,\;\;b_2=-\frac{2 c^2 (2 \gamma -\alpha -1)}{\alpha -1},\; \omega=\omega.$$
Therefore, proceeding as in the previous cases, the set of solutions for this family is provided by
\begin{equation} \label{S-15}
	\begin{split}
		u_{15}(x,t) & = \frac{2 c^2 (6 \gamma ^2-\gamma  \alpha -\gamma -\alpha ^2-2 \alpha -1)}{(\alpha -1) (9 \gamma -7 \alpha -7)} \tanh^{2}(cx-\omega t)+b_1 \coth(cx-\omega t)\\ & -\frac{2 c^2 (2 \gamma -\alpha -1)}{\alpha -1}\coth^{2}(cx-\omega t).
	\end{split}
\end{equation}
\noindent {\bf Family 16:} For $\alpha\neq\pm 1$, $c\neq 0$ and $2 \gamma -\alpha -1\neq 0$:
$$ a_0=a_0, \; \; a_1=a_1,\;\; a_2=0,\;\; b_1=0,\;\; b_2=b_2,\; \omega=\frac{a_0 c\alpha+3\gamma b_{2}^2}{(1-\alpha^2)(2 \gamma -\alpha -1)}.$$
Therefore, proceeding as in the previous cases, the set of solutions for this family is provided by
\begin{equation} \label{S-16}
	u_{16}(x,t)  =a_0+ a_1\tanh(cx+\omega t)+b_2\coth^{2}(cx-\omega t).
\end{equation}
\noindent {\bf Family 17:} For $\alpha\neq 1$, $c\neq 0$ and $ \gamma -\alpha -1\neq 0$:
$$ a_0=0, \; \; a_1=0,\;\; a_2=\frac{2 c^2 (9 \gamma ^2-\alpha ^2-2 \alpha-1)}{9 (\alpha-1) (\gamma -\alpha-1)},\;
b_1=b_1,\;\;b_2=-\frac{2 c^2 (3 \gamma -\alpha -1)}{3 (\alpha -1)},\; \omega=\omega.$$
Therefore, proceeding as in the previous cases, the set of solutions for this family is provided by
\begin{equation} \label{S-17}
	u_{17}(x,t)  =\frac{2 c^2 (9 \gamma ^2-\alpha ^2-2 \alpha-1)}{9 (\alpha-1) (\gamma -\alpha-1)}\tanh^{2}(cx+\omega t)+b_1\coth(cx-\omega t) -\frac{2 c^2 (3 \gamma -\alpha -1)}{3 (\alpha -1)} \coth^{2}(cx-\omega t).
\end{equation}
\noindent {\bf Family 18:} For $\alpha\neq  1$, $c\neq 0$ and $ \gamma \neq 0$:
$$ a_0=-\frac{a_2 (3 a_2 \alpha+10 \gamma  c^2 +2 c^2)}{2 c^2 \gamma }, \; \; a_1=0,\;\; a_2=a_2,\;
b_1=b_1,\;\;b_2=-\frac{2 c^2 (2 \gamma -\alpha -1)}{\alpha -1},\;\;  \omega=\omega.$$
Therefore, proceeding as in the previous cases, the set of solutions for this family is provided by
\begin{equation} \label{S-18}
	u_{18}(x,t)  =-\frac{a_2 (3 a_2 \alpha+10 \gamma  c^2 +2 c^2)}{2 c^2 \gamma }+a_2 \tanh^{2}(cx+\omega t)+b_1\coth(cx-\omega t) -\frac{2 c^2 (2 \gamma -\alpha -1)}{\alpha -1} \coth^{2}(cx-\omega t).
\end{equation}
\noindent {\bf Family 19:} For $\alpha\neq  1$, $ \gamma \neq 0$, $c\neq 0$ and $2a_1\neq b_1$:
$$ a_0=-\frac{a_1 (a_1^2 \alpha ^2-2 a_1^2 \alpha +a_1^2-8 \gamma ^2 c^4+4 \gamma  c^4 \alpha +4 \gamma  c^4)}{2 c^2 \gamma  (\alpha-1) \left(2 a_1-b_1\right)}, \; \; a_1=a_1,\;\; a_2=0,$$
$$ b_1=b_1,\;\;b_2=-\frac{2 c^2 (2 \gamma -\alpha -1)}{\alpha -1},\;\; \omega=\omega.$$
Therefore, proceeding as in the previous cases, the set of solutions for this family is provided by
\begin{equation} \label{S-19}
	\begin{split}
		u_{19}(x,t) & =-\frac{a_1 (a_1^2 \alpha ^2-2 a_1^2 \alpha +a_1^2-8 \gamma ^2 c^4+4 \gamma  c^4 \alpha +4 \gamma  c^4)}{2 c^2 \gamma  (\alpha-1) \left(2 a_1-b_1\right)}+a_1 \tanh(cx-\omega t)+b_1 \coth(cx-\omega t)\\ & -\frac{2 c^2 (2 \gamma -\alpha -1)}{\alpha -1}\coth^{2}(cx-\omega t).
	\end{split}
\end{equation}
\noindent {\bf Family 20:} For $\alpha\neq\pm 1$, $c\neq 0$ and $\gamma\neq 0$:
$$ a_0=0, \; \; a_1=a_1,\;\; a_2=0,\;
b_1=\frac{2c \sqrt{a_1(3 \gamma -\alpha -1)}}{\sqrt{3}\sqrt{\alpha -1}},\;\;b_2=-\frac{ c^4 (3 \gamma -5\alpha-1)}{ \gamma^2(\alpha^2 -1)},\;\;  \omega=\omega.$$
Therefore, proceeding as in the previous cases, the set of solutions for this family is provided by
\begin{equation} \label{S-20}
	u_{20}(x,t)  =a_1 \tanh(cx+\omega t) + \frac{2c \sqrt{a_1(3 \gamma -\alpha -1)}}{\sqrt{3}\sqrt{\alpha -1}}\coth(cx-\omega t) -\frac{ c^4 (3 \gamma -5\alpha-1)}{ \gamma^2(\alpha^2 -1)} \coth^{2}(cx-\omega t).
\end{equation}
\noindent {\bf Family 21:} For $ \gamma \neq 0$ and $c\neq 0$:
$$ a_0=0, \; \; a_1=0,\;\; a_2=a_2,\;\;
b_1=\pm\frac{a_2 \sqrt{\alpha+1}}{\sqrt{\gamma }},\;\;b_2=0,\;\;  \omega=\omega.$$
Therefore, proceeding as in the previous cases, the set of solutions for this family is provided by
\begin{equation} \label{S-21}
	u_{21}(x,t)  =a_2 \tanh^{2}(cx+\omega t)\pm\frac{a_2 \sqrt{\alpha+1}}{\sqrt{\gamma }}\coth(cx-\omega t).
\end{equation}
\noindent {\bf Family 22:} For $\alpha\neq \pm 1$, $c\neq 0$ and $ \gamma \neq 0$:
$$ a_0=\frac{3 \alpha ^3-2 \gamma  c^2}{c \gamma  (\alpha^2 -1)}, \; \; a_1=0,\;\; a_2=-a_0,\;\;
b_1=0,\; b_2=0,\;\;  \omega=\frac{3a_0 \alpha}{4\gamma c^2 (\alpha-1)}.$$
Therefore, proceeding as in the previous cases, the set of solutions for this family is provided by
\begin{equation} \label{S-22}
	u_{22}(x,t)  =\frac{3 \alpha ^3-2 \gamma  c^2}{c \gamma  (\alpha^2 -1)}-\frac{3 \alpha ^3-2 \gamma  c^2}{ c \gamma  (\alpha^2 -1)}\tanh^{2}(cx+\omega t).
\end{equation}
\noindent {\bf Family 23:} For $\alpha\neq  1$, $c\neq 0$ and $ \gamma \neq 0$:
$$ a_0=\frac{3a_1^4 \alpha ^2-2 \gamma  c^4}{4 c^2a_1 \gamma  (\alpha -1)}, \; \; a_1=a_1\neq 0,\;\; a_2=0,\;\;
b_1=0,\; b_2=-\frac{2 c^2 (2 \gamma -\alpha-1)}{\alpha -1},\;\;  \omega=\omega.$$
Therefore, proceeding as in the previous cases, the set of solutions for this family is provided by
\begin{equation} \label{S-23}
	u_{23}(x,t)  =\frac{3a_1^4 \alpha ^2-2 \gamma  c^4}{4 c^2a_1 \gamma  (\alpha -1)}+a_1\tanh(cx+\omega t)-\frac{2 c^2 (2 \gamma -\alpha-1)}{\alpha -1} \coth^{2}(cx-\omega t).
\end{equation}
\noindent {\bf Family 24:} For $\alpha\neq \pm 1$, $\gamma\neq 0$, $c\neq 0$ and  $2 \gamma -\alpha -1\neq 0$:
$$ a_0=a_0, \; \; a_1=a_1\;\; a_2=-\frac{10 c^2 (2 \gamma -\alpha -1)}{3 (\alpha -1)},\;\;
b_1=\frac{3 a_1 \sqrt{3a_0 \gamma -\alpha -1}}{\sqrt{2c \gamma }},$$ $$b_2=-\frac{2 a_{1}c^2 (2 \gamma -\alpha -1)}{\alpha^2 -1},\;\;  \omega=\frac{a_0 c^3 \alpha}{ 3(2 \gamma -\alpha -1)^2}.$$
Therefore, proceeding as in the previous cases, the set of solutions for this family is provided by
\begin{equation} \label{S-24}
	\begin{split}
		u_{24}(x,t) & =a_0 +a_1 \tanh(cx-\omega t)-\frac{10 c^2 (2 \gamma -\alpha -1)}{3 (\alpha -1)} \tanh^{2}(cx-\omega t)\\&+\frac{3 a_1 \sqrt{3a_0 \gamma -\alpha -1}}{\sqrt{2c \gamma }} \coth(cx-\omega t)-\frac{2 a_{1}c^2 (2 \gamma -\alpha -1)}{\alpha^2 -1}\coth^{2}(cx-\omega t).
	\end{split}
\end{equation}
As we can see in \cite{Ref1,Ref2,Ref3,Sal,Sal1} and its references, the  technique proposed here has been effectively applied by various authors to solve problems involving shallow water waves.
\section{Graphical presentation of solutions and discussion}\label{Gr}
\noindent In this part, we will exhibit in 3D and 2D some of the solutions revealed in the previous section in order to study the evolution of traveling waves and their dependence on $\alpha$, $\gamma$, and other relevant parameters.

\vspace{0.1in}

\begin{figure}[h!]
	\begin{center}
		\includegraphics[scale=0.75]{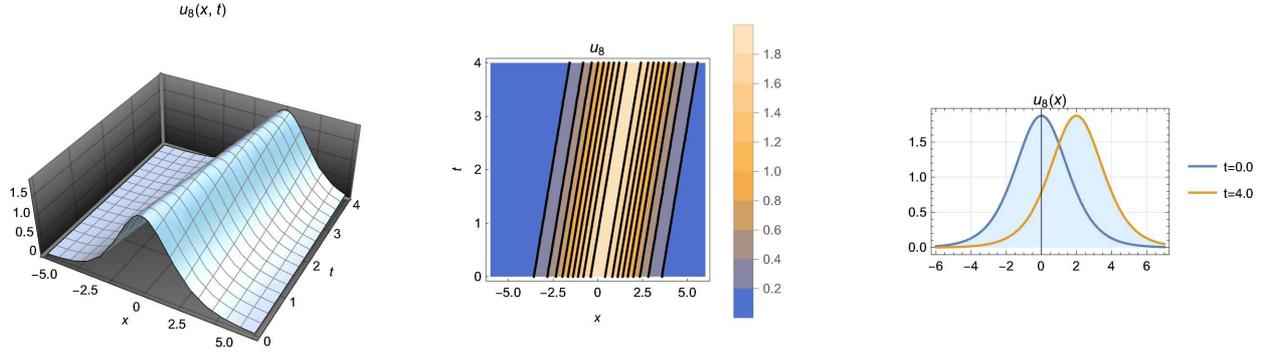}
	\end{center}
	\caption{Solution profile $u_8(x,t)$ in the interval $-5.0\leq x\leq 5.0$, for the parameters selected in Example 1 (left). Wavefront contour plot (center), and 2D plot of traveling wave solution $u_8(x,t)$ for $t = 0.0$ and $t = 4.0$ (right).}\label{Fig-1}
\end{figure}

\begin{figure}[h!]
	\begin{center}
		\includegraphics[scale=0.75]{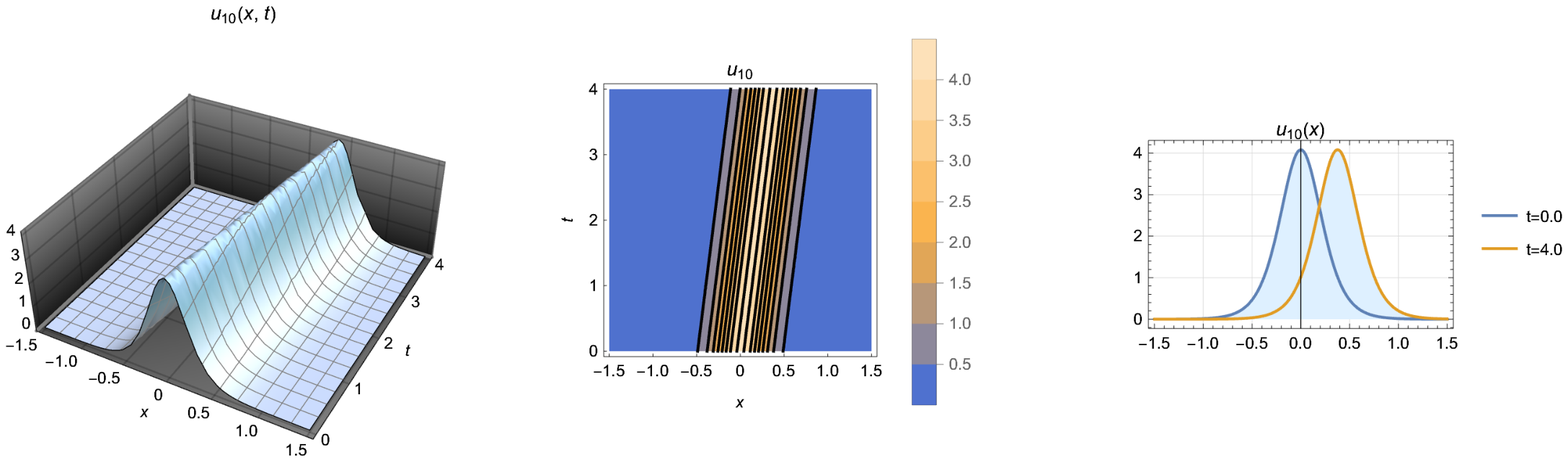}
	\end{center}
	\caption{Solution profile $u_{10}(x,t)$ in the interval $-1.5\leq x\leq 1.5$, for the parameters selected in Example 2 (left). Wavefront contour plot (center), and  2D plot of traveling wave solution $u_{10}(x,t)$ for $t = 0.0$ and $t = 4.0$ (right).}\label{Fig-2}
\end{figure}

\begin{figure}[h!]
	\begin{center}
		\includegraphics[scale=0.75]{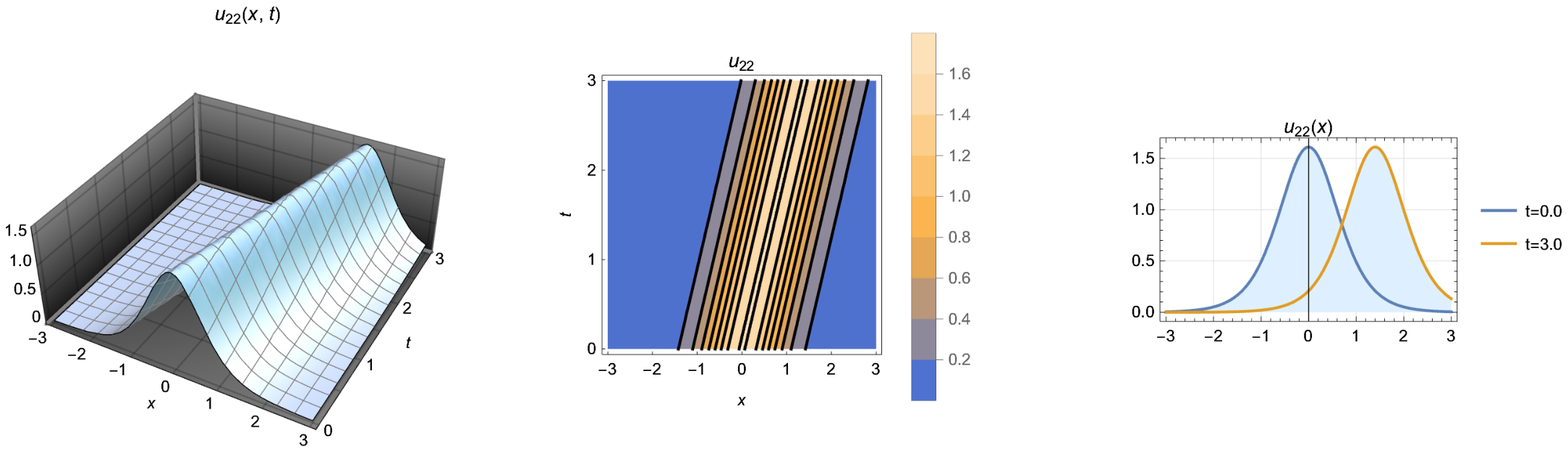}
	\end{center}
	\caption{Solution profile $u_{22}(x,t)$ in the interval $-3.0\leq x\leq 3.0$, for the parameters selected in Example 3 (left). Wavefront contour plot (center), and 2D plot of traveling wave solution $u_{22}(x,t)$ for $t = 0.0$ and $t = 3.0$ (right).}\label{Fig-3}
\end{figure}

\begin{figure}[h!]
	\begin{center}
		\includegraphics[scale=0.75]{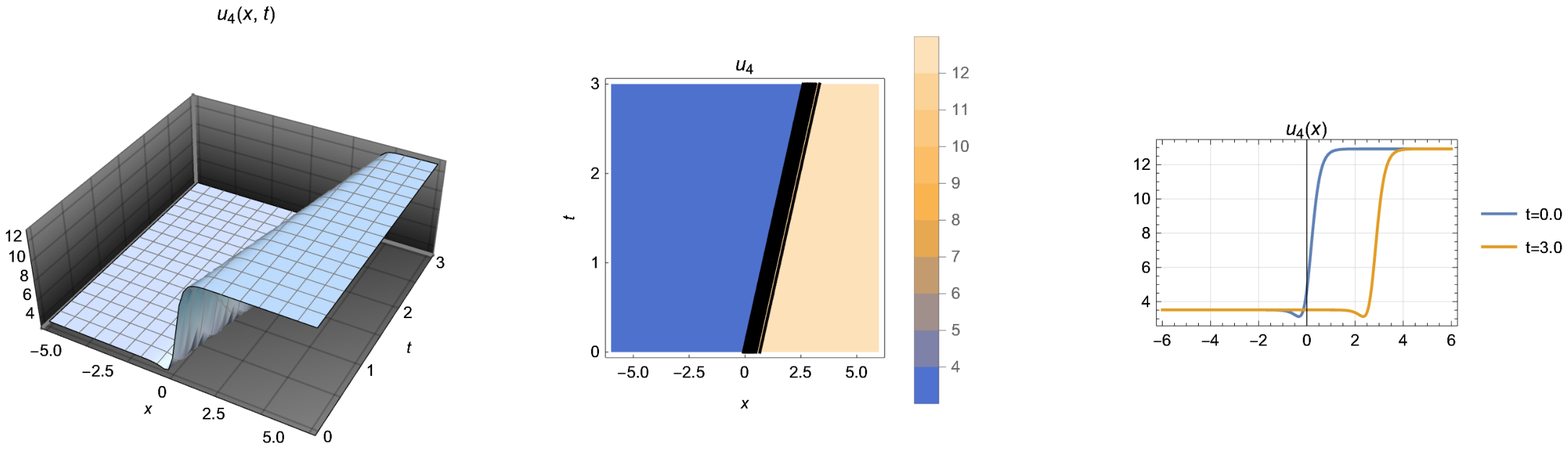}
	\end{center}
	\caption{Solution profile $u_{4}(x,t)$ in the interval $-5.0\leq x\leq 5.0$, for the parameters selected in Example 4 (left). Wavefront contour plot (center), and 2D plot of traveling wave solution $u_{4}(x,t)$ for $t = 0.0$ and $t = 3.0$ (right).}\label{Fig-4}
\end{figure}

\begin{figure}[h!]
	\begin{center}
		\includegraphics[scale=0.75]{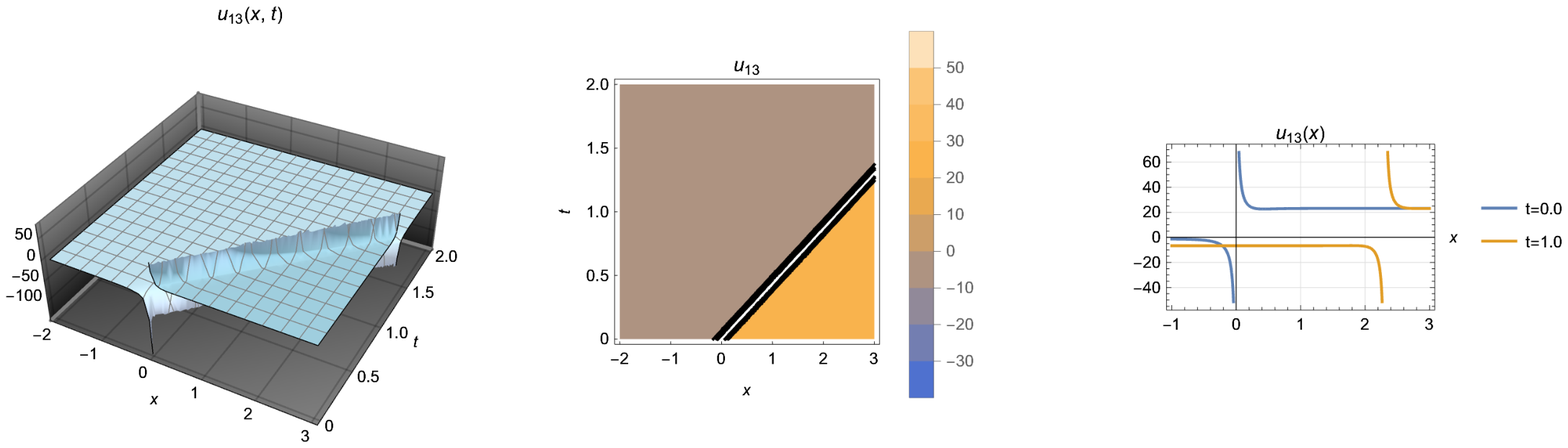}
	\end{center}
	\caption{Solution profile $u_{13}(x,t)$ in the interval $-3.0\leq x\leq 3.0$, for the parameters selected in Example 5 (left). Wavefront contour plot (center), and 2D plot of traveling wave solution $u_{13}(x,t)$ for $t = 0.0$ and $t = 1.0$ (right).}\label{Fig-5}
\end{figure}

\begin{figure}[h!]
	\begin{center}
		\includegraphics[scale=0.75]{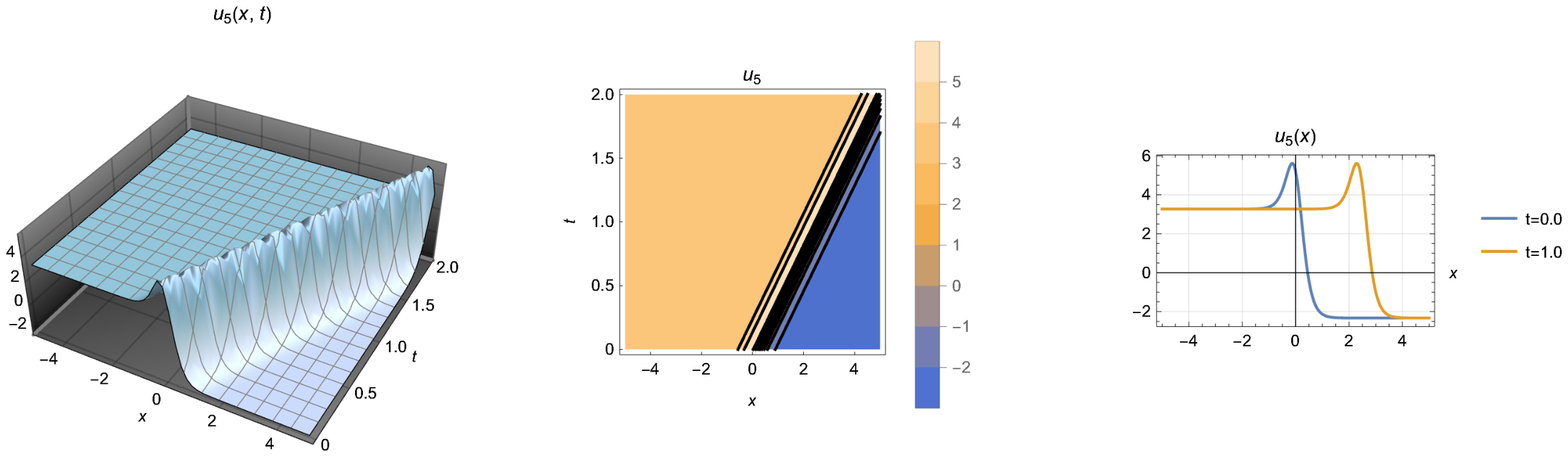}
	\end{center}
	\caption{Solution profile $u_{5}(x,t)$ in the interval $-5.0\leq x\leq 5.0$, for the parameters selected in Example 6 (left). Wavefront contour plot (center), and 2D plot of traveling wave solution $u_{5}(x,t)$ for $t = 0.0$ and $t = 1.0$ (right).}\label{Fig-6}
\end{figure}

\begin{figure}[h!]
	\begin{center}
		\includegraphics[scale=0.75]{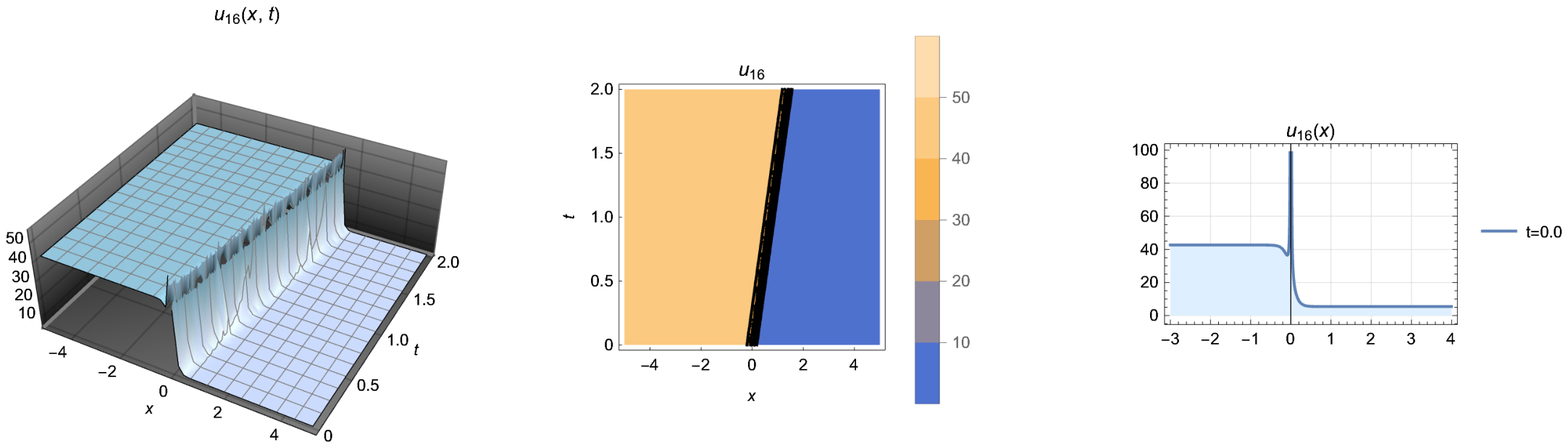}
	\end{center}
	\caption{Solution profile $u_{16}(x,t)$ in the interval $-5.0\leq x\leq 5.0$, for the parameters selected in Example 16 (left). Wavefront contour plot (center), and 2D plot of traveling wave solution $u_{16}(x,t)$ for $t = 0.0$ (right).}\label{Fig-7}
\end{figure}

\noindent {\bf Example 1.} To illustrate, let us consider the Family 8 with the parameters $\alpha=1$, $\gamma=1$, and $c=0.5$. Which allows us to determine that $a_2=-1.87$ and $a_0=1.875$. In addition to the free coefficients we choose them as $a_1=0$, $b_1=0$, $b_2=0$ and $\omega=0.25$. Figure \ref{Fig-1} shows the 3D and 2D  bell-shaped solution $u_{8}(x,t)$ for these parameters. \\
{\bf Example 2.} To illustrate, let us consider the Family 10 with the parameters $\alpha=-1$, $\gamma=-1$, and $c=3.5$. Which allows us to determine that $a_2=-4.08$ and $a_0=4.08$. In addition to the free coefficients we choose them as $a_1=0$, $b_1=0$, $b_2=0$ and $\omega=0.33$. Figure \ref{Fig-2} shows the 3D and 2D  bell-shaped solution $u_{10}(x,t)$ for these parameters. \\
{\bf Example 3.} To illustrate, let us consider the Family 22 with the parameters $\alpha=4$, $\gamma=2$, and $c=1.2$. Which allows us to determine that $a_0=1.61$, $a_2=-1.61$ and $\omega=0.56$.  Figure \ref{Fig-3} shows the 3D and 2D  bell-shaped solution $u_{22}(x,t)$ for these parameters. \\
{\bf Example 4.} To illustrate, let us consider the Family 4 with the parameters $\alpha=2$, $\gamma=1.5$, and $c=2.4$. Which allows us to determine that $a_2=3.52$, $b_2=0$ and $\omega=2.14$. In addition to the free coefficients we choose them as $a_0=4.7$ and $a_1=4.7$. Figure \ref{Fig-4} shows the 3D and 2D  kink-shaped solution $u_{4}(x,t)$ for these parameters. \\
{\bf Example 5.} To illustrate, let us consider the Family 13 with the parameters $\alpha=3.01$, $\gamma=2.2$, and $c=2.6$. Which allows us to determine that $a_0=8.22$ and $b_1=6.51$.  In addition to the free coefficients we choose them as $a_1=5.7$, $a_2=2.7$ and $\omega=6.0$. Figure \ref{Fig-5} shows the 3D and 2D  singular traveling wave solution $u_{13}(x,t)$ for these parameters. \\
{\bf Example 6.} To illustrate, let us consider the Family 5 with the parameters $\alpha=3.4$, $\gamma=2.2$, and $c=-2.4$. Which allows us to determine that $a_2=-4.73$ and $b_2=0$.  In addition to the free coefficients we choose them as $a_0=5.2$, $a_1=2.8$ and $\omega=-5.8$. Figure \ref{Fig-6} shows the 3D and 2D  anti-kink-shaped solution $u_{5}(x,t)$ for these parameters. \\
{\bf Example 7.} To illustrate, let us consider the Family 16 with the parameters $\alpha=5.5$, $\gamma=2.0$, and $c=6.0$. Which allows us to determine that $\omega=4.15$.  In addition to the free coefficients we choose them as $a_0=23$, $a_1=-18.6$, $a_2=0$, $b_1=0$ and $b_{2}=1.05$. Figure \ref{Fig-7} shows the 3D and 2D  singular-shaped solution $u_{16}(x,t)$ for these parameters. \\
\noindent Since $\alpha=1$ and $\gamma=1$, Family 8 studied in Example 1 represents a set of solutions for the SIdV Eq. (\ref{K1}), whereas Family 10 studied in Example 2 represents a set of solutions for the KdV Eq. (\ref{K0}). In addition, the Family 22 examined in Example 3 is a solution set for the gsiaKdV equation (\ref{Eq-1}), because $\alpha\neq \pm1$ and $\gamma\neq0$.  These three examples illustrate that the proposed technique confirms what has been stated by previous researchers \cite{KdV3,Re-3,Re-1,Re-5}, namely that Eqs. (\ref{K0}), (\ref{K1}) and (\ref{Eq-1}) have solutions of the bell-shaped type (\ref{K01}). Finally, the families considered in Examples 4, 5, 6, and 7 are solution sets for the gsiaKdV equation (\ref{Eq-1}), because $\alpha \neq\pm 1$ and $\gamma\neq 0$, and it is shown that Eq. (\ref{Eq-1}) admits traveling wave solutions of the kink, anti-kink, and singular anti-kink varieties \cite{Re-1}.

\section{ Conclusions}\label{C}
\noindent The gsiaKdV equation is a generalization of both the KdV and SIdV equations; the mathematical model includes  the term $\big(3(1-\alpha)u+(1+\alpha)\frac{u_{xx}}{u}\big)$ can be seen as an advecting velocity. In this paper, we study the gsiaKdV equation for the first time using the tanh-coth method, and the results revealed that this generalization shares bell-shaped solutions with both the KdV equation and the SIdV equation, as reported previously by other authors, also has other kinds of solutions that could be of interest in the study of shallow wave motion over the ocean. We believe that the research of novel traveling wave solutions will shed light on new sorts of applications of this equation in applied mathematics and engineering, as well as depict the specific behaviour of atmospheric phenomena caused by shallow ocean waves.  In future study, other exact solution methods may be used to the studied equation.

\section*{Data availability statement}
\noindent Our article have no associated data.

\vspace{0.2in}

\section*{Declaration of Competing Interest} 
\noindent The authors declare that they have no known competing financial interests or personal relationships that could have appeared to influence the work reported in this paper.

\section*{Funding Information}
\noindent The authors declare that no funds, grants, or other support were received during the preparation of this manuscript.



%
%

\end{document}